\documentclass[10pt]{article}
\usepackage{color}
\usepackage{amsfonts}
\usepackage{amssymb}
\usepackage{amsthm}
\usepackage{amsmath}
\usepackage{graphics}
\usepackage{graphicx}
\usepackage{a4wide}
\usepackage{enumerate}

\begin{document}

\newcommand\independent{\protect\mathpalette{\protect\independenT}{\perp}}
\def\independenT#1#2{\mathrel{\rlap{$#1#2$}\mkern2mu{#1#2}}}
\newcommand{\<}{\langle}
\renewcommand{\>}{\rangle}

\theoremstyle{definition}
\newtheorem{Def}{Definition}

\theoremstyle{remark}
\newtheorem*{Rem}{Remarks}
\newtheorem{Ex}[Def]{Example}

\theoremstyle{plain}
\newtheorem{Lem}[Def]{Lemma}
\newtheorem{Prop}[Def]{Proposition}
\newtheorem{Thm}[Def]{Theorem}
\newtheorem{Cor}[Def]{Corollary}
\newcommand{\eqL}[1]{\stackrel{#1}{=}}
\newcommand{\leqL}[1]{\stackrel{#1}{\leq}}
\newcommand{\geqL}[1]{\stackrel{#1}{\geq}}

\title{Information-theoretic inference of common ancestors}
\author{ Bastian Steudel$^1$
\;and Nihat Ay$^{1,2}$\\
$ $\\
\small{\texttt{steudel@mis.mpg.de}}\\
$ $\\
$^1$MPI for Mathematics in the Sciences, Leipzig, Germany\\
$^2$Santa Fe Institute, New Mexico, USA
}
\maketitle

\begin{abstract}
A directed acyclic graph (DAG) partially represents the conditional independence structure among observations of a system if the local Markov condition holds, that is, if every variable is independent of its non-descendants given its parents. In general, there is a whole class of DAGs that represents a given set of conditional independence relations. We are interested in properties of this class that can be derived from observations of a subsystem only. To this end, we prove an information theoretic inequality that allows for the inference of common ancestors of observed parts in \emph{any} DAG representing some unknown larger system. More explicitly, we show that a large amount of dependence in terms of mutual information among the observations implies the existence of a common ancestor that distributes this information. Within the causal interpretation of DAGs our result can be seen as a quantitative extension of Reichenbach's Principle of Common Cause to more than two variables.\\
Our conclusions are valid also for non-probabilistic observations such as binary strings, since we state the proof for an axiomatized notion of  `mutual information' that includes the stochastic as well as the algorithmic version.
\end{abstract}

\section{Introduction}

Causal relations among components $X_1,\ldots, X_n$ of a system are commonly modeled in terms of a directed acyclic graph (DAG) in which there is an edge $X_i\rightarrow X_j$ whenever $X_i$ is a direct cause of $X_j$. Further, it is usually assumed that information about the causal structure can be obtained through interventions in the system. However, there are situations in which interventions are not feasible (too expensive, unethical or physically impossible) and one faces the problem to infer causal relations from \emph{observational} data only. To this end, postulates linking observations to the underlying causal structure have been employed, one of the most fundamental being the \emph{causal Markov condition}\,\cite{pearlCausality,spirtes}. It connects the underlying causal structure to conditional independencies among the observations. Explicitly it states that every observation is independent of its non-effects given its direct causes. It formalizes the intuition, that the only relevant components of a system for a given observation are its direct causes.\\
In terms of DAGs, the causal Markov condition states that a DAG can only be a valid causal model of a system if every node is independent of its non-descendants given its parents. The graph is then said to fulfill the \emph{local Markov condition}\,\cite{lauritzen96}. Consider for example the causal hypothesis $X\rightarrow Y \leftarrow Z$ on three observations $X, Y$ and $Z$. Assuming the causal Markov condition, the hypothesis implies that $X$ and $Z$ are independent. Violation of this independence then allows one to exclude this causal hypothesis. But note that in general there are many DAGs that fulfill the local Markov condition with respect to a given set of conditional independence relations. For example, all three DAGs $X\rightarrow Y\rightarrow Z$, $X\leftarrow Y\rightarrow Z$ and $X\leftarrow Y\leftarrow Z$ encode that $X$ is independent of $Z$ given $Y$ and it can not be decided from information on conditional independences alone, which is the true causal model. Nevertheless, properties that are shared by all valid DAGs (e.g. an edge between $X$ and $Y$ in the example) provide information about the underlying causal structure.\\
The causal Markov condition is only expected to hold for a given set of observations if all relevant components of a system have been observed, that is if  there are no confounders (causes of more than two observations that have not been measured). It can then be proven by assuming a functional model of causality\,\cite{pearlCausality,dominikK,bastianCOLT}. As an example, consider the observations $X_1,\ldots,X_n$ to be jointly distributed random variables. In this case, the causal Markov condition can be derived for a given DAG on $X_1,\ldots,X_n$ from two assumptions: $(1)$  every variable $X_i$ is a \emph{deterministic function} of its parents and an independent (possibly unobserved) noise variable $N_i$ and $(2)$ the noise variables $N_i$ are jointly independent. However, in this paper we assume that our observations provide only partial knowledge about  a system and ask for structural properties common to all DAGs that represent the independencies of some \emph{larger} set of elements.\\
To motivate our result, assume first that our observation consists of only two jointly distributed random variables $X_1$ and $X_2$ which are stochastically dependent. Reichenbach\,\cite{reichenbach} postulated already in 1956 that the dependence of $X_1$ and $X_2$ needs to be explained by (at least) one of the following cases: $X_1$ is a cause of $X_2$, or $X_2$ is a cause of $X_1$, or there exists a common cause of $X_1$ and $X_2$. This link between dependence and the underlying causal structure is known as \emph{Reichenbach's principle of common cause}. It is easily seen that by assuming  $X_1$ and $X_2$ to be part of some unknown larger system whose causal structure is described by a DAG $G$, then the causal Markov condition for $G$ implies the principle of common cause. Moreover, we can subsume all three cases of the principle if we formally allow a node to be an ancestor of itself and arrive at\\

\noindent {\bf Common cause principle.}
{\em If two observations $X_1$ and $X_2$ are dependent, then they must have a common ancestor in \emph{any} DAG modeling some possibly larger system.\/}\\

\noindent Our main result is an information theoretic inequality that enables us to generalize this principle to more than two variables. It leads to the\\

\noindent {\bf Extended common cause principle (informal version).}
{\em Consider $n$ observations $X_1, \dots, X_n$, and a number $c$, $1\leq
c\leq n$. If the dependence of the observations exceeds a bound that depends on $c$, then in \emph{any} DAG modeling some possibly larger system there exist $c$ nodes out of $X_1,\ldots,X_n$ that have a common ancestor.}\\

\begin{figure}
\centering
  \includegraphics[width=0.5\textwidth]{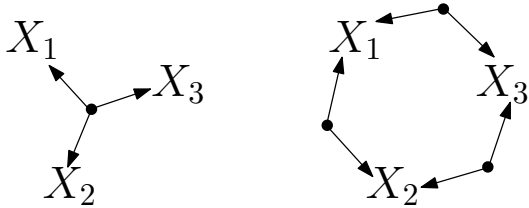}\\
  \caption{Two causal hypothesis for which the causal Markov condition does not imply conditional independencies among the observations $X_1,X_2$ and $X_3$. Thus they can not be distinguished using qualitative criteria like the common cause principle (unobserved variables are indicated as dots). However, the model on the right can be excluded if the dependence among the $X_i$ exceeds a certain bound. }\label{figCRIntro}
\end{figure}

\noindent Thus, structural information can be obtained by exploiting
the degree of dependence on the subsystem and we would
like to emphasize that, in contrast to the original common cause
principle, the above criterion provides means to distinguish among
cases with the same independence structure of the observed
variables. This is illustrated in Figure \ref{figCRIntro}.

\noindent Above, the extended common cause principle is stated without making explicit the kind of observations we consider and how dependence is quantified. In the main case we have in mind, the observations are jointly distributed random variables and dependence is  quantified by the mutual information\,\cite{coverThomas} function. Then the extended common cause principle (Theorem \ref{thmDisc}) relates stochastic dependence to a property of all Bayesian networks that include the observations.\\
However, the result holds for more general observations (such as binary strings) and for more general notions of mutual information (such as algorithmic mutual information\,\cite{algStat}). Therefore we introduce an 'axiomatized' version of mutual information in the following Section and describe how it can be connected to a DAG. Then, in Section \ref{secAn} we prove a theorem on the decomposition of information about subsets of a DAG out of which the extended common cause principle then follows as a corollary. Apart from a larger area of applicability, we think that an abstract proof based on an axiomatized notion of information better illustrates that the result is independent of the notion of 'probability'. It only relies on the basic properties of (stochastic) mutual information (see Definition \ref{defMutual}). Finally, in Section \ref{secApp} we describe the result in more detail within different contexts and relate it to the notion of redundancy and synergy that was introduced in the area of neural information processing. 

\section{General mutual information and DAGs}

\begin{figure}
\centering
  \includegraphics[width=0.8\textwidth]{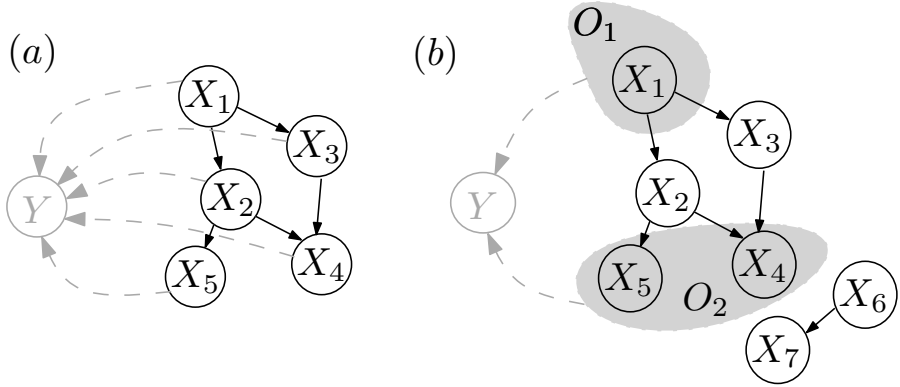}\\
  \caption{The graph in $(a)$ shows a DAG on nodes $X_1,\ldots,X_5$ whose observation is modeled by a leaf node $Y$ (e.g. a  noisy measurement). Figure $(b)$ shows a DAG-model of observed elements $O_1=\{X_1\}$ and $O_2=\{X_4,X_5\}$.}\label{figY}
\end{figure}

Before introducing a general notion of mutual information, let us describe how it is connected to a DAG in the stochastic setting. Assume we are given an observation of $n$ discrete random variables $X_1,\ldots,X_n$ in terms of their joint probability distribution $p(X_1,\ldots,X_n)$. Write $[n]=\{1,\ldots,n\}$ and for a subset $S\subseteq [n]$ let $X_S$ be the random variable associated with the tuple $(X_i)_{i\in S}$. Assume further, that a  directed acyclic graph (DAG) $G$ is associated with the nodes $X_1,\ldots,X_n$, that fulfills the local Markov condition\,\cite{lauritzen96}: for all $i, (1\leq i \leq n)$
\begin{equation}\label{relInd}
X_i \independent X_{nd_i}\quad |\; X_{pa_i}\,,
\end{equation}
where $nd_i$ and $pa_i$ denote the subset of indices corresponding to the non-descendants and to the parents of $X_i$ in $G$. The tuple $(G,p(X_{[n]}))$ is called a Bayesian net\,\cite{pearlBelief} and the conditional independence relations imply the factorization of the joint probability distribution
$$
p(x_1,\ldots,x_n) = \prod_{i\in[n]} p(x_i|x_{pa_i})\,,
$$
where small letters $x_i$ stand for values taken by the random variables $X_i$.
From this factorization it follows that the joint information measured in terms of Shannon entropy\,\cite{coverThomas} decomposes into a sum of individual conditional entropies
\begin{equation}\label{decEnt}
H(X_1,\ldots,X_n) = \sum_{i=1}^n H(X_i\,|\,X_{pa_i})\,.
\end{equation}
Shannon entropy can be considered as absolute measure of information. However, in many cases only a notion of information \emph{relative} to another observation may be available. For example, in the case of continuous random variables, Shannon entropy can be negative and hence may not be a good measure of the information. Therefore we would like formulate our results based on a relative measure, such as mutual information, which, moreover, induces a notion of independence in a natural way. This can be achieved by introducing a specially designated variable $Y$ relative to which information will be quantified. $Y$ can for example be thought of as providing a noisy measurement of the $X_{[n]}$ (Fig. \ref{figY} $(a)$).
Then, with respect to a joint probability distribution $p(Y,X_{[n]})$  we can transform the decomposition of entropies into a decomposition of mutual information\,\cite{coverThomas}
\begin{equation}\label{decMut}
I(Y\,:\,X_{[n]}) \geq \sum_{i=1}^n I(Y\,:\,X_i\;|X_{pa_i})\,.
\end{equation}
For a proof and a condition for equality see Lemma \ref{lemDecMut} below. In the case of discrete variables, Shannon entropy $H(X_i)$ can be seen as mutual information of $X_i$ with a copy of itself: $H(X_i)=I(X_i:X_i)$. Therefore we can always choose $p(Y|X_{[n]})$ such that $Y=X_{[n]}$ and the decomposition of entropies in (\ref{decEnt}) is recovered. We are interested in decompositions as in (\ref{decEnt}) and (\ref{decMut}), since their violation allows us to exclude possible DAG structures.\\
\noindent However, note that the above relations are not yet very useful, since they require, through the assumption of the local Markov condition, that we have observed all relevant variables of a system. Before we relax this assumption in the next section we introduce mutual information measures on general observations.
\begin{Def}[measure of mutual information]$ $\label{defMutual}\\
Given a finite set of elements $\mathcal{O}$ a measure of mutual information on $\mathcal{O}$ is a three-argument function on the power set 
$$
I: 2^{\mathcal{O}}\times 2^{\mathcal{O}} \times 2^{\mathcal{O}} \rightarrow \mathbb{R},\quad (A,B,C) \rightarrow I(A:B\,|C)
$$
such that for \emph{disjoint} sets $A,B,C,D\subseteq 2^{\mathcal{O}}$ it holds:
\begin{eqnarray*}
I(A:\emptyset) &=& 0\quad \text{(normalization)}\\
I(A:B\,|C) &\geq &  0\quad \text{(non-negativity)}\\
I(A:B\,|C) &=& I(B:A\,|C)\quad \text{(symmetry)}\\
I(A:\,(B\cup C)\,|\,D) &=&  I(A:B\,|\,C\cup D) + I(A:C\,|\,D)\quad\text{ (chain rule).}   
\end{eqnarray*}
We say $A$ is independent of $B$ given $C$ and write $(A\independent B\,|C)$ iff $I(A:B\,|C)=0$. Further we will generally omit the empty set as a third argument and substitute the union by a comma, hence we write $I(A:B)$ instead of $I(A:B\,|\emptyset)$ and $I(A:B,C)$ instead of $I(A:B\cup C)$.
\end{Def}
Of course, mutual information of discrete as well as of continuous random variables is included in the above definition. Further, in Section \ref{secStrings} we will discuss a recently developed theory of causal inference\,\cite{dominikK} based on \emph{algorithmic} mutual information of binary strings\footnote{Mutual information of composed quantum systems satisfies the definition as well, because it can be defined in formal analogy to classical information theory if Shannon entropy is replaced by von Neumann entropy of a quantum state. The properties of mutual information stated above have been used to single out quantum physics from a whole class of no-signaling theories\,\cite{infoCausQuant}.}. We now state two properties of mutual information that we need later on.
\begin{Lem}[properties of mutual information]$ $\label{lemPropMut}\\
Let $I$ be a measure of mutual information on a set of elements $\mathcal{O}$. Then
\begin{enumerate}
\item[$(i)$] (data processing inequality) For three disjoint sets $A,B,C\subseteq \mathcal{O}$
$$
I(A:C\,|B) = 0 \quad\Longrightarrow\quad I(A:B)\geq I(A:C)\,.
$$
\item[$(ii)$] (increase through conditioning on independent sets)\\For three disjoint sets $A,B,C\subseteq \mathcal{O}$
\begin{equation}\label{ineqCond}
I(A:C\,|B)=0 \quad\Longrightarrow\quad I(Y:A\,|B) \leq I(Y:A\,|B,C)\,,
\end{equation}
where $Y$ is an arbitrary set $Y\subseteq \mathcal{O}$ disjoint from the rest. Further, the difference is given by $I(A:C\,|B,Y)$. 
\end{enumerate}
\end{Lem}
\noindent Proof: $(i)$ Using the chain rule two times
\begin{eqnarray*}
I(A:B) &=& I(A:B) + I(A:C\,|B) = I(A:B,C)\\
&=& I(A:C) + I(A:B|C) \geq I(A:C),
\end{eqnarray*}
where the last inequality follows from non-negativity of $I$. To prove $(ii)$ we again use the chain rule 
\begin{eqnarray*}
I(Y:A\,|B) - I(Y:A\,|B,C) &=& I(Y:A\,|B) - I(Y,C:A\,|B) + I(A:C|B)\\
&=& -I(A:C\,|B,Y) \leq 0\,.
\end{eqnarray*}
$\Box$

\noindent As in the stochastic setting, we can connect a DAG to the conditional independence relation that is induced by mutual information: we say that a DAG on a given set of observations \emph{fulfills the local Markov condition} if every node is independent of its non-descendants given its parents. Furthermore, we show in Appendix \ref{apDsep} that the induced independence relations are sufficiently nice, in the sense that they satisfy the semi-graphoid axioms \cite{dawidCI}. This is useful because it implies that a DAG that fulfills the local Markov condition is an efficient partial representation of the conditional independence structure. Namely, conditional independence relations can be read off the graph with the help of a criterion called d-separation\,\cite{pearlCausality} (see Appendix \ref{apDsep} for details).\\

\noindent We conclude with a general formulation of the decomposition of mutual information that we already described in the probabilistic case. 
\begin{Lem}[decomposition of mutual information]$ $\label{lemDecMut}\\
Let $I$ be a measure of mutual information on elements $O_{[n]}=\{O_1,\ldots,O_n\}$ and $Y$. Further let $G$ be a DAG with node set $O_{[n]}$ that fulfills the local Markov condition. Then
\begin{equation}\label{inDecMut}
I(Y:O_{[n]}) \geq \sum_{i=1}^n I(Y:O_i\,|O_{pa_i})
\end{equation}
with equality if conditioning on $Y$ does preserve the independences of the local Markov condition: that is for all $i$
\begin{equation}\label{inLem}
O_i \independent O_{nd_i}\;|(O_{pa_i},Y)\,.
\end{equation}
\end{Lem}
Proof: Assume the $O_i$ are ordered topologically with respect to $G$. The proof is by induction on $n$. The lemma is trivially true if $n=1$ with equality. Assume that it holds for $k-1<n$. It is easy to see that the graph $G_{k}$ with nodes $O_{[k]}$ that is obtained from $G$ by deleting all but the first $k$ nodes fulfills the local Markov condition with respect to $O_{[k]}$. By the chain rule 
$$
I(Y:O_{[k]}) = I(Y:O_{[k-1]}) + I(Y: O_{k}\,|O_{[k-1]})
$$
and we are left to show that $I(Y: O_{k}\,|O_{[k-1]})\geq I(Y: O_{k}\,|O_{pa_{k}})$. Since the local Markov condition holds, we have $O_{k} \independent O_{[k-1]\backslash pa_{k}}\,|O_{pa_{k}}$ and the inequality follows by applying (\ref{ineqCond}). Further, by property $(ii)$ of the previous Lemma, equality holds if for every $k$: $O_{k} \independent O_{[k-1]\backslash pa_{k}}\,|\,(O_{pa_{k}},Y)$ which is implied by (\ref{inLem}). $\Box$\\

\noindent In the next section we derive a similar inequality in the case in which  only the mutual information of $Y$ with a subset of the nodes $O_{[n]}$ is known.

\section{Partial information about a system}\label{secAn}

We have shown that the information about elements of a system described by a DAG  decomposes if the graph fulfills the local Markov condition. In this section we derive a similar decomposition in cases where not all elements of a system have been observed. This decomposition will of course depend on specific properties of $G$ and, in turn, enable us to exclude certain DAGs as models of the \emph{total} system whenever we observe a violation of such a decomposition.\\
More precisely, we are interested in properties of the class of DAG-models of a set of observations that we define as follows (see Figure \ref{figY} $(b)$).
\begin{Def}[DAG-model of observations]$ $\label{defDAGmodel}\\
An \emph{observation} of elements $O_{[n]}=\{O_1,\ldots,O_n\}$ with respect to a reference object $Y$ and mutual information measure $I$ is given by the values of  $I(Y:O_{S})$ for every subset $S\subseteq [n]$.\\
A DAG $G$ with nodes $\mathcal{X}$ together with a measure of mutual information $I_G$ on $\mathcal{X}$ is a \emph{DAG-model} of an observation, if the following holds
\begin{itemize}
\item[$(i)$] each observation $O_i$ is a subset of the nodes $\mathcal{X}$ of $G$.
\item[$(ii)$] $G$ fulfills the local Markov condition with respect to $I_G$
\item[$(iii)$] $I_G$ is an extension of $I$, that is $I_G(Y:O_S) = I(Y:O_S)$ for all $S\subseteq [n]$.
\item[$(iv)$] $Y$ is a leaf node (no descendants) of $G$\,.
\end{itemize}
\end{Def}
\noindent The first three conditions state that, given the causal Markov condition, $G$ is a valid hypothesis on the causal relations among components of some larger system including the $O_{[n]}$ that is consistent with the observed mutual information values. Condition $(iv)$ is merely a technical condition due to the special role of $Y$ as an observation of the $O_{[n]}$ external to the system.\\
As an example, if the $O_i$ and $Y$ are random variables with joint distribution $p(O_{[n]},Y)$, a DAG-model $G$ with nodes $\mathcal{X}$ is given by the graph structure of a Bayesian net with joint distribution $p(\mathcal{X})$, such that the marginal on $O_{[n]}$ and $Y$ equals $p(O_{[n]},Y)$. Moreover, if $Y$ is a copy of $O_{[n]}$ then an observation in our sense is given by the values of the Shannon entropy $H(O_{S})$ for every subset $S\subseteq [n]$.\\
The general question posed in this paper can then be formulated as follows: What can be learned from an observation given by the values $I(Y:O_{S})$ about the class of DAG-models?\\
As a first step we present a property of mutual information about independent elements.
\begin{Lem}[submodularity of $I$]$ $\label{lemSubMod}\\
If the $O_i$ are \emph{mutually independent}, that is $I(O_i : O_{[n]\backslash i})=0$ for all $i$, then the function $[n] \supseteq S\rightarrow -I(Y\,:\,O_S)$ is submodular, that is, for two sets $S,T\subseteq [n]$ 
$$
I(Y\,:\,O_S) + I(Y\,:\,O_T) \leq I(Y\,:\,O_{S\cup T}) + I(Y\,:\,O_{S\cap T})\,.
$$
\end{Lem}
\noindent Proof: For two subsets $S,T\subseteq [n]$ write $S' = S\backslash (S\cap T)$ and $T' = T\backslash (S\cap T)$. Using the chain rule we have
\begin{eqnarray*}
I(Y:O_{S\cup T}) + I(Y:O_{S\cap T}) &=& I(Y\,:\,O_{S}) + I(Y:O_{T'}\,|O_S) +I(Y:O_{S\cap T})\\
&\geq & I(Y\,:\,O_{S}) + I(Y:O_{T'}\,|O_{S\cap T}) +I(Y:O_{S\cap T})\\
&=& I(Y\,:\,O_S) + I(Y:O_T)\,,
\end{eqnarray*}
where the inequality follows from property (\ref{ineqCond}) of mutual information. $\Box$\\

\begin{figure}
\centering
  \includegraphics[width=0.8\textwidth]{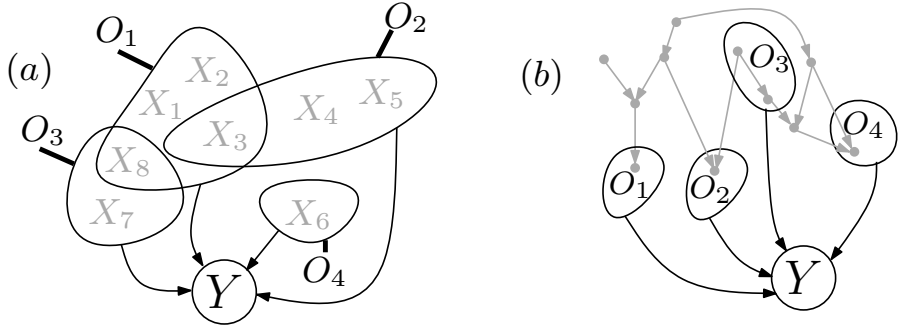}\\
  \caption{$(a)$ shows four subsets $O_1,\ldots,O_4$ of independent elements $X_1,\ldots,X_8$ `observed by' $Y$. Note that the intersection of three sets $O_i$ is empty, hence $d_i\leq 2$ for all $i=1,\ldots,4$ in Proposition \ref{propInd} and therefore $I(Y:O_{[4]}) \geq \frac{1}{2}\sum_{i=1}^4 I(Y:O_i)$. $(b)$ shows a DAG-model in gray. The observed elements $O_1,\ldots,O_4$ are subsets of its nodes. One can check that the DAG does not imply any conditional independencies among the $O_i$ (e.g. with the help of the $d$-separation criterion, see Appendix \ref{apDsep}). Nevertheless, there is no common ancestor of all four observations ( $\cap_{i=1}^4 an(O_i)=\emptyset$). Since $Y$ only depends on the $O_i$, inequality (\ref{ineqInd2}) of Theorem \ref{thmAn} implies $I(Y:O_{[4]}) \geq \frac{1}{3}\sum_{i=1}^4 I(Y:O_i)$.}\label{figAn}
\end{figure}

\noindent Hence, a violation of submodularity allows one to reject mutual independence among the $O_{i}$ and therefore to exclude the DAG that does not have any edges from the class of possible DAG-models (the local Markov condition would imply mutual independence).\\
We now broaden the applicability of the above Lemma based on a result for submodular functions from \cite{madiman08}: We assume that there are unknown objects $\mathcal{X}=\{X_1,\ldots,X_r\}$ which are mutually independent and that the observed elements $O_i\subseteq \mathcal{X}$ will be subsets of them (see Figure \ref{figAn} $(a)$).
In contrast to the previous lemma it is not required anymore, that the $O_i$ are mutually independent themselves. It turns out, that the way the information about the $O_i$ decomposes allows for the inference of intersections among the sets $O_i$, namely
\begin{Prop}[decomposition of information about sets of independent elements]$ $\label{propInd}\\
Let $\mathcal{X}=\{X_1,\ldots,X_r\}$ be mutually independent objects, that is $I(X_j:X_{[r]\backslash j})=0$ for all $j$. Let $O_{[n]}=\{O_1,\ldots,O_n\}$, where each $O_i\subseteq \mathcal{X}$ is a non-empty subset of $\mathcal{X}$. For every $i \in [n]$ let $d_i$ be maximal such that $O_i$ has non-empty intersection with $d_i-1$ sets out of $O_{[n]}$ distinct from $O_i$.
Then the information about the $O_{[n]}$ can be bounded from below by
\begin{equation}\label{ineqProp}
I(Y:\,O_{[n]}) \geq \sum_{i=1}^{n} \frac{1}{d_i}I(Y:O_i)\,.
\end{equation}
\end{Prop}
\noindent For an illustration see Figure \ref{figAn}(a). Even though the proposition is actually a corollary of the following theorem, its proof is given in appendix \ref{propIndProof} since it is, unlike the theorem, independent of graph theoretic notions\,.\\
As a trivial example consider the case where $O_1=O_2=O\subseteq\mathcal{X}$ are identical subsets. Then $d_1=d_2=2$ and
$$
I(Y:O) = \frac{1}{2}I(Y:O_1) + \frac{1}{2}I(Y:O_2),
$$
hence equality holds in (\ref{ineqProp}). In general, if there is an element in $O_i$, that is also in $k-1$ different sets $O_j$, then $d_i\geq k$ and we account for this redundancy in dividing the single information $I(Y:O_i)$ by at least $k$.\\
Independent elements can always be modeled as root nodes of a DAG. The following theorem, that is our main result, generalizes the proposition by connecting the information about observations $O_i$ to the intersection structure of associated ancestral sets. For a given DAG $G$, a set of nodes $A$ is called ancestral, if for every edge $v\rightarrow w$ in $G$ such that $w$ is in $A$, also $v$ is in $A$. Further, for a subset of nodes $S$, we denote by $an(S)$ be the smallest ancestral set that contains $S$. Elements of $an(S)$ will be called ancestors of $S$.\\

\begin{Thm}[decomposition of ancestral information]$ $\label{thmAn}\\
Let $G$ be a DAG-model of an observation of elements $O_{[n]}=\{O_1,\ldots,O_n\}$. For every $i$ let $d_i$ be the maximal number such that the intersection of $an(O_i)$ with $d_i-1$ distinct sets $an(O_{i_1}),\ldots,an(O_{i_{d-1}})$ is non-empty. Then the information about all ancestors of $O_{[n]}$ can be bounded from below by
\begin{equation}\label{ineqInd}
I(Y:\,an(O_{[n]})) \geq \sum_{i=1}^{n} \frac{1}{d_i}I(Y:an(O_i)) \geq  \sum_{i=1}^{n} \frac{1}{d_i}I(Y:O_i)\,.
\end{equation}
Furthermore, if $Y$ only depends on whole system $\mathcal{X}$ through the $O_{[n]}$, that is 
\begin{equation}\label{indYY}
Y\independent \mathcal{X}\backslash (O_{[n]}\cup \{Y\})\;|\,O_{[n]}\,
\end{equation}
we obtain an inequality containing only known values of mutual information
\begin{equation}\label{ineqInd2}
I(Y:\,O_{[n]}) \geq  \sum_{i=1}^{n} \frac{1}{d_i}I(Y:O_i)\,.
\end{equation}
\end{Thm}
\noindent The proof is given in Appendix \ref{thmAnProof} and an example is illustrated in Figure \ref{figAn}(b). If all quantities except the structural parameters $d_i$ are known, inequality (\ref{ineqInd2}) can be used to obtain information about the intersection structure among the $O_i$ that is encoded in the $d_i$ provided that the independence assumption (\ref{indYY}) holds. Even if (\ref{indYY}) does not hold but information on an upper bound of $I(Y:an(O_{[n]}))$ is available (e.g. in terms of the entropy of $Y$) information about the intersection structure may be obtained from $(\ref{ineqInd})$. The following corollary additionally provides a bound on the minimum information about ancestral sets. 
\begin{Cor}[inference of common ancestors, local version]$ $\label{corLoc}\\
Given an observation of elements $O_{[n]}=\{O_1,\ldots,O_n\}$, assume that for natural numbers $\mathbf{c}=(c_1,\ldots,c_n)$ with $(1\leq c_i \leq n-1)$ we observe
\begin{equation}\label{ineqCor}
 \epsilon_{\mathbf{c}} := \sum_{i=1}^{n} \frac{1}{c_i}I(Y:O_i) - I(Y:\,an(O_{[n]})) > 0.
\end{equation}
Let $G$ be an arbitrary DAG-model of the observation. For every $O_i$, let $A_{c_i+1}$ be the set of common ancestors in $G$ of $O_i$ and at least $c_i$ elements of $O_{[n]}$ different from $O_i$. Then the joint information about all common ancestors can be bounded from below by 
$$
I\big(Y\,:\, \cup_{i=1}^n A_{c_i+1}\big) \;\geq\; \big(\sum_{i=1}^n \frac{1}{c_i} - 1\big)^{-1} \epsilon_{\mathbf{c}} \;>\; 0\,.
$$
In particular, for an index $i\in[n]$ we must have $A_{c_i+1}\neq \emptyset$, hence  there exists a common ancestor of $O_i$ and at least $c_i$ elements of $O_{[n]}$ different from $O_i$.
\end{Cor} 
\noindent The proof is given in Appendix \ref{apCorProof}. Theorem \ref{thmAn} and its corollary are our most general results but due to ease of interpretation we illustrate them in the next section only in the speciale case in which all $c_i$ are equal (Cor. \ref{corRed}) to obtain a lower bound on the information about all common ancestors of at least $c+1$ elements $O_i$.\\
To conclude this section, we ask what is the maximum amount of information that one can expect to obtain about the intersection structure of ancestral sets of a DAG-model of an observations. The main requirement for a DAG-model $G$ is, that it fulfills the local Markov condition with respect to some larger set $\mathcal{X}$ of elements. This will remain true if we add nodes and arbitrary edges in a way that $G$ remains acyclic. Therefore, if $G$ contains a common ancestor of $c$ elements we can always construct a DAG-model $G'$ that contains a common ancestor of more than $c$ elements (e.g. the DAG-model on the right hand side of Fig. \ref{figCRIntro} can be transformed in the one on the left hand side). We conclude that  without adding minimality requirements for the DAG-models (such as the causal faithfulness assumption\,\cite{spirtes}) only assertions on ancestors of a \emph{minimal} number of nodes can be made.

\section{Structural implications of redundancy and synergy}\label{secApp}

The results of the last section can be related to the notions of redundancy and synergy. In the context of neuronal information processing, it has been proposed\,\cite{stillNetworkInfo} to capture the redundancy and synergy of elements $O_{[n]}=\{O_{1},\ldots,O_{n}\}$ with respect to another element $Y$ using the function
\begin{equation}\label{eqRedundancy}
r(Y) := \sum_{i=1}^n I(Y\,:\,O_{i}) -  I(Y\,:\,O_{[n]})\,,
\end{equation}
where $I$ is a measure of mutual information. Thus $r$ relates information that $Y$ has about the single elements to information about the whole set.\\
If the sum of informations about the single $O_i$ is larger than the information about whole set ($r(Y)>0$), the $O_{[n]}$ are said to be \emph{redundant} with respect to $Y$. This may be the case if $Y$ `contains' information that is shared by multiple $O_i$. In general, if the $O_i$ do not share any information, that is, if they are mutually independent, then they can not be redundant with respect to any $Y$ (this follows from Lemma \ref{lemSubMod}).\\
On the other hand, if the information of $Y$ about the whole set of elements is larger than about its single elements ($r(Y)<0$), the $O_{[n]}$ are called \emph{synergistic} with respect to $Y$. This may for example be the case if $Y$ is generated through a function $Y=f(O_1,\ldots,O_n)$ and  the function value contains little information about each argument (as is the case for the parity function, see below). If, instead, $Y$ is a copy of the $O_{[n]}$, then $r(Y)\geq 0$ and thus the $O_{[n]}$ are not synergetic with respect to $Y$.\\%\footnote{If $r(Y)=0$, then conditioning on $Y$ does not change the mutual information among the $O_{[n]}$, in particular it preserves the independence structure (Proposition \ref{propSyn} below).}\\
To connect our results to the introduced notion of redundancy and synergy, we introduce the following version of $r$ parametrized by a parameter $c\in \{1,\ldots,n\}$ 
\begin{equation}\label{eqRedFine}
r_{c}(Y) := \frac{1}{c}\sum_{i=1}^n  I(Y\,:\,O_{i}) -  I(Y\,:\,O_{[n]})\,.
\end{equation}
Intuitively, if $r_{c}(Y)>0$ for large $c$, then the $O_i$ are highly redundant with respect to $Y$. Corollary \ref{corLoc} of the last section implies that high redundancy implies common ancestors of many $O_i$.
\begin{Cor}[redundancy explained structurally]$ $\label{corRed}\\
Let an observation of elements $O_{[n]}=\{O_1,\ldots,O_n\}$ be given by the values of $I(Y:O_{S})$ for any subset $S\subseteq [n]$. If $r_{\mathbf{c}}(Y) > 0$, then in any DAG-model of the observation in which $Y$ only depends on $\mathcal{X}$ through $O_{[n]}$\,\footnote{We formulate the independence assumption as $Y \independent \tilde{\mathcal{X}}\,|O_{[n]}$, where $\tilde{\mathcal{X}}$ denotes all nodes of the DAG-model different from the nodes in $O_{[n]}$ and $Y$. Note that this assumption does not hold in the  original context in which $r$ has been introduced. There, $Y$ is the observation of a stimulus that is presented to some neuronal system and the $O_{i}$ represent the responses of (areas of) neurons to this stimulus.}, there exists a common ancestor of at least $c+1$ elements of $O_{[n]}$.
\end{Cor}
\noindent In the following two subsections we discuss this result in more detail for the cases in which the observed elements are discrete random variables and binary strings. 

\subsection{Common ancestors of discrete random variables}

Let $X_{[n]}=\{X_1,\ldots,X_n\}$ and $Y$ be discrete random variables with joint distribution $p(X_{[n]},Y)$ and let $I$ denote the usual measure of mutual information given by the Kullback-Leibler divergence of $p$ from its factorized distribution\,\cite{coverThomas}. If $Y=X_{[n]}$ is a copy of the $X_{[n]}$ then $I(Y:X_{[n]}) = H(X_{[n]})$, where $H$ denotes the Shannon entropy. In this case the redundancy $r_1(X_{[n]})$ is equal to the multi-information\,\cite{studenyMulti} of the $X_{[n]}$. Moreover $r_c$ gives rise to a parametrized version of multi-information 
$$
I_c(X_1,\ldots,X_n) :=  \sum_{i=1}^n \frac{1}{c} H(X_{i}) - H(X_{[n]})\,,
$$
and from Corollary $\ref{corLoc}$ we obtain
\begin{Thm}[lower bound on entropy of common ancestors]$ $\label{thmDisc}\\
Let $X_{[n]}$ be jointly distributed discrete random variables. If $I_c(X_{[n]}) > 0$, then, in \emph{any} Bayesian net containing the $X_{[n]}$, there exists a common ancestor of strictly more than $c$ variables out of the $X_{[n]}$. Moreover, the entropy of the set $A_{c+1}$ of all common ancestors of more than $c$ variables is lower bounded by 
$$
H(A_{c+1}) \geq \frac{c}{n-c} I_c(X_{[n]})\,.
$$
\end{Thm}

\begin{figure}
\centering
  \includegraphics[width=0.3\textwidth]{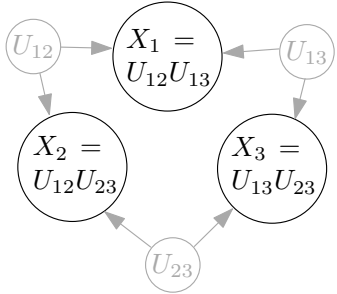}\\
  \caption{ The figure illustrates that higher order interaction among observed random variables can be explained by a Bayesian net in which only common ancestors of two variables exist. More precisely, all random variables are assumed to be binary with values in $\{-1,1\}$ and the unobserved common ancestors $U_{ij}$ are mutually independent and uniformly distributed. Further the value of each observation $X_i$ is obtained the product of the values of its two ancestors. Then the resulting marginal distribution $p(X_1,X_2,X_3)$ is of higher order interaction: it is related to the parity function $p(X_1=x_1,X_2=x_2,X_3=x_3) = \frac{1}{4}$ if $x_1x_2x_3=1$ and zero otherwise.}\label{figXOR}
\end{figure}

\noindent We continue with some remarks to illustrate the theorem:\\
$\mathbf{(a)}$ Setting $c=1$, the theorem states that, up to a factor $1/(n-1)$, the multi-information $I_1$ is a lower bound on the entropy of common ancestors of more than two variables. In particular, if $I_1(X_{[n]})>0$ any Bayesian net containing the $X_{[n]}$ must have at least an edge.\\
$\mathbf{(b)}$ Conversely, the entropy of common ancestors of all the elements $X_1,\ldots,X_n$ is lower bounded by $(n-1)I_{n-1}(X_{[n]})$. This bound is not trivial whenever $I_{n-1}(X_{[n]})>0$, which is for example the case if the $X_i$ are only slightly disturbed copies of some not necessarily observed random variable (see example below).\\
$\mathbf{(c)}$ We emphasize that the inferred common ancestors can be among the elements $X_i$ themselves. Unobserved common ancestors can only be inferred by postulating assumptions on the causal influences \emph{among} the $X_i$. If, for example, all the $X_i$ were measured simultaneously, a direct causal influence among the $X_i$ can be excluded and any dependence or redundancy has to be attributed to unobserved common ancestors.\\
$\mathbf{(d)}$ Finally note that $I_c>0$ is only a sufficient, but not a necessary condition for the existence of common ancestors. However, we know that the information theoretic information provided by $I_c$ is used in the theorem in an optimal way. By this we mean that we can construct distributions $p(X_{[n]})$, such that $I_c(X_{[n]})=0$ for a given $c$ and no common ancestors of $c+1$ nodes have to exist.\\
We conclude this section with two examples:\\
\textbf{Example (three variables):} Let $X_1,X_2$ and $X_3$ be three binary variables, each with maximal entropy $H(X_i)=\log 2$. Then $I_2(X_1,X_2,X_3)>0$ iff the joint entropy $H(X_1,X_2,X_3)$ is strictly less than $\frac{3}{2}\log 2$. In this case, there must exist a common ancestor of all three variables in any Bayesian net that contains them. In particular, any Bayesian net corresponding to the DAG on the right hand side of Figure \ref{figCRIntro} can be excluded as a model.\\
\textbf{Example (synchrony and interaction among random variables):} Let $X_1=X_2=\cdots=X_n$ be identical random variables with non-vanishing entropy $h$. Then in particular $I_{n-1}(X_{[n]}) = (n-1)^{-1}h > 0$ and we can conclude that there has to exist a common ancestor of all $n$ nodes in any Bayesian net that contains them.\\
In contrast to the synchronized case, let $X_1,X_2,\ldots,X_n$ be binary random variables taking values in $\{-1,1\}$ and assume that the joint distribution is of pure $n$-interaction\footnote{This terminology is
motivated by the general framework of interaction spaces proposed
and investigated by Darroch et. al.\,\cite{speed} and used by Amari\,\cite{amari} within information geometry.}, that is for some $\beta \neq 0$ it has the form 
$$
p_\beta(x_1,\ldots,x_n) := \frac{1}{Z_\beta}\exp(\beta x_1x_2\cdots x_n), 
$$
where $Z$ is a normalization constant. It can be shown that there exists a Bayesian net including the $X_{[n]}$, in which common ancestors of \emph{at most} two variables exist. This is illustrated in Figure \ref{figXOR} for three variables and in the limiting case $\beta=\infty$ in which each $X_i$ is uniformly distributed and $X_1=X_2 \cdot X_3$.
We found it somewhat surprising that, contrary to synchronization, higher order interaction among observations does not require common ancestors of many variables.\\

\subsection{Common ancestors in string manipulation processes}\label{secStrings}

In some situations it is not convenient or straightforward to summarize an observation in terms of a joint probability distribution of random variables. Consider for example cases in which the data comes from repeated observations under varying conditions (e.g. time series). A related situation is given if the number of samples is low. Janzing and Schoelkopf\,\cite{dominikK} argue that causal inference in these situations still should be possible, provided that the observations are sufficiently complex. To this end, they developed a framework for causal inference from \emph{single} observations that we describe now briefly. Assume we have observed two objects $A$ and $B$ in nature (e.g. two carpets) and we encoded these observations into binary strings $a$ and $b$. If the descriptions of the observations in terms of the strings $a$ and $b$ are sufficiently complex and sufficiently similar (e.g. the same pattern on the carpets) one would expect an explanation of this similarity in terms of a mechanism that relates these two strings in nature (are the carpets produced by the same company?). It is necessary that the descriptions are sufficiently complex, as an example of\,\cite{dominikK} illustrates: assume the two observed strings are equal to the first hundred digits of the binary expansion of $\pi$, hence they can be generated independently by a simple rule. If this is the case, the similarity of the two strings would not be considered as strong evidence for the existence of a causal link. To exclude such cases, Kolmogorov complexity\,\cite{bookKC} $K(s)$ of a string $s$ has been used as measure of complexity. It is defined as the length of the shortest program that prints out $s$ on a universal (prefix-free) Turing machine. With this definition, strings that can be generated using a simple rule, such as the constant string $s=0\cdots0$ or the first $n$ digits of the binary expansion of $\pi$ are considered simple, whereas it can be shown that a random string of length $n$ is complex with high probability. Kolmogorov complexity can be transformed into a function on sets of strings by choosing a suitable concatenation function $\<\cdot,\cdot\>$, such that $K(s_1,\ldots,s_n) = K(\<s_1,\<s_2,\ldots,\<s_{n-1},s_n\>\ldots\>)$.\\
The \emph{algorithmic} mutual information\,\cite{algStat} of two strings $a$ and $b$ is then equal to the sum of the lengths of the shortest programs that generate each string separately minus the length of the shortest program that generates the strings $a$ and $b$:
$$
I(a:b) \stackrel{+}{=} K(a)+K(b)-K(a,b)\,,
$$
where $\stackrel{+}{=}$ stands for equality up to an additive constant that depends on the choice of the universal Turing machine. Analog to Reichenbach's principle of common cause, \cite{dominikK} postulates a causal relation among $a$ and $b$ whenever $I(a:b)$ is large, which is the case if the complexities of the strings are large and both strings together can be generated by a much shorter program than the programs that describe them separately.\\
In formal analogy to the probabilistic case, algorithmic mutual information can be extended to a conditional version defined for sets of strings $A,B,C\subseteq \{s_1,\ldots,s_n\}$ as
$$
I(A:B\,|C) \stackrel{+}{=} K(A\cup C) + K(B\cup C) - K(A\cup B\cup C) - K(C)\,.
$$
Intuitively, $I(A:B\,|C)$ is the mutual information between the strings of $A$ and the strings of $B$ if a shortest program that prints the strings in $C$ has been provided as an additional input. Based on this notion of condition mutual information the causal Markov condition can be formulated in the algorithmic setting. It can be proven\,\cite{dominikK} to hold for a directed acyclic graph $G$ on strings $s_1,\ldots,s_n$ if every $s_i$ can be computed by a simple program on a universal Turing machine from its parents and an additional string $n_i$ such that the $n_i$ are mutually independent. Without going into the details we sum up by stating that DAGs on strings can be given a causal interpretation and it is therefore interesting to infer properties of the class of possible DAGs that represent the algorithmic conditional independence relations.\\
In the algorithmic setting, our result can be stated as follows
\begin{Thm}[inference of common ancestors of strings]$ $\\
Let $O_{[n]}=\{s_1,\ldots,s_n\}$ be a set of  binary strings. If for a number $c, (1\leq c \leq n-1)$
$$
\frac{1}{c} \sum_{i=1}^n K(s_i) - K(s_1,\ldots,s_n) \stackrel{+}{\geq} 0,
$$
then there must exist a common ancestor of at least $c+1$ strings out of $O_{[n]}$ in any DAG-model of the $O_{[n]}$.\footnote{Here $\stackrel{+}{\geq}$ means up to an additive constant dependent only on the choice of a universal Turing machine, on $c$ and on $n$.}
\end{Thm}
\noindent Proof: As described, algorithmic mutual information is an information measure in our sense only up to an additive constant depending on the choice of the universal Turing machine. However, one can check that in this case, the decomposition of mutual information (Theorem \ref{thmAn}) holds up to an additive constant that depends additionally on the number of strings $n$ and the chosen parameter $c$. The result on Kolmogorov complexities follows by choosing $Y=(s_1,\ldots,s_n)$, since $K(s_i) \stackrel{+}{=} I(Y:s_i)$. $\Box$\\

\noindent Thus, highly redundant strings require a common ancestor in any DAG-model. Since the Kolmogorov complexity of a string $s$ is uncomputable, we have argued in recent work\,\cite{bastianCOLT}, that it can be substituted by a measure of complexity in terms of the length of a compressed version of $s$ with respect to a chosen compression scheme (instead of a universal Turing machine) and the above result should still hold approximately.

\subsection{Structural implications from synergy?}\label{secSyn}

We saw that large redundancy implies common ancestors of many elements and we may wonder whether structural information can be obtained from synergy in a similar way. This seems not to be possible, since synergy is related to more fine-grained information (information about the mechanisms) as the following example shows: Assume the observations $O_{[n]}$ are mutually independent. Then \emph{any} DAG is a valid DAG-model since the local Markov condition will always be satisfied. We also now that $r(Y)\leq 0$, but it turns out that the amount of synergy crucially depends on the way that $Y$ has processed the information of the $O_{[n]}$ (and therefore not on a structural property among the $O_{[n]}$ themselves). To see this, let the observations $O_i$ be binary  random variables which are mutual independent and distributed uniformly, such that
$$
p(O_{[n]}) = \prod_{i=1}^n p(O_i) \quad \text{ and } p(O_i = 1) = p(O_i = 0) = 1/2\,.
$$
Further let $Y = (O_i\oplus O_j)_{i<j}$ be a function of the observations (addition is modulo $2$). Then the $O_{[n]}$ are highly synergetic with respect to $Y$, that is $r_1(Y) = -(n-1)\log 2$. On the other hand, if $Y=O_1\oplus\cdots\oplus O_n$, then $r_1(Y) = -\log 2$ only.\\
Nevertheless, it is an easy observation that synergy with respect to $Y$ can be related to an increase of redundancy after conditioning on $Y$. Since $I(\cdot\,|Y)$ is a measure of mutual information as well, we define a conditioned version of $r$ in a canonical way as
$$
r_c(Z|Y) = \frac{1}{c}\sum_{i=1}^n I(Z:O_i\,|Y) - I(Z : O_{[n]}|Y)\,,
$$
with respect to some observation $Z$. If $I$ can be evaluated on non-disjoint subsets, that is, if we can choose $Z=O_{[n]}$, we have the following
\begin{Prop}[synergy from increased redundancy induced by conditioning]$ $\label{propSyn}\\
Let $O_{[n]}=\{O_{1},\ldots,O_{n}\}$ and $Y$  be arbitrary elements on which a mutual information function $I$ is defined. Then 
$$
r_c(Y) = r_c(O_{[n]}) - r_c(O_{[n]}|Y)\,,
$$
hence if conditioning on $Y$ increases the redundancy of $O_{[n]}$ with respect to itself, then $r_c(Y)<0$ and the $O_{[n]}$ are synergetic with respect to $Y$.
\end{Prop} 
\noindent Proof: Using the chain rule, we derive
$$
r_c(O_{[n]})-r_c(O_{[n]}|Y) = r_c(Y) - r_c(Y|O_{[n]}) = r_c(Y)\,,
$$
where the last equality follows because $r_c(Y|O_{[n]}) = 0$.
$\Box$\\
Continuing the example of binary random variables above, mutual independence of the $O_{[n]}$ is equivalent to $r_1(O_{[n]})=0$ and therefore, using the proposition $r_1(Y) = -r_1(O_{[n]}|Y)$. Thus, if $Y=O_1\oplus\cdots\oplus O_{[n]}$, 
$$
r_1(Y) = -r_1(O_{[n]}|Y) = H(O_{[n]}|Y)-\sum_{i=1}^n H(O_i|Y) = -\log 2\,,
$$
as already noted above.

\section{Discussion}

Based on a generalized notion of mutual information, we proved an inequality describing the decomposition of information about a whole set into the sum of information about its parts. The decomposition depended on a structural property, namely the existence of common ancestors in a DAG. We connected the result to the notions of redundancy and synergy and concluded that large redundancy implies the existence of common ancestors in any DAG-model. Specialized to the case of discrete random variables, this means that large stochastic dependence in terms of multi-information needs to be explained through a common ancestor (in a Bayesian net) acting as a broadcaster of information.\\
Much work has been done already that examined the restrictions that are imposed on observations by graphical models that include latent variables. Pearl\,\cite{pearlCausality, pearlTestability} already investigated constraints imposed by the special instrumental variable model. Also Darroch et al.\,\cite{speed} and recently Sullivant et. al \cite{sullivantGauss} looked at linear Gaussian graphical models and determined constraints in terms of the entries on the covariance matrix describing the data (tetrad constraints). Further, methods of algebraic statistics were applied\,(e.g. \cite{ricco07}) to derive constraints that are induced by latent variable models directly on the level of probabilities. In general this does not seem to be an easy task due to the large number of variables involved and information theoretic quantities allow for relatively easy derivations of `macroscopic' constraints (see also \cite{nihat2}).\\
Finally, we think that the general methodology of connecting concepts such as \emph{synergy} and \emph{redundancy} of observations to properties of the class of possible DAG-models is interesting, especially in the light of their causal interpretation.

\appendix

\section{Semi-graphoid axioms and $d$-separation}\label{apDsep}

Consider the conditional independence relation that is induced by an information measure on a set of objects ($A\independent B|C\;\Leftrightarrow\;I(A:B\,|C)=0$). Then 
\begin{Lem}[general independence satisfies semi-graphoid axioms]$ $\label{lemGraphoid}\\
The relation of (conditional) independence induced by an independence measure $I$ on elements $\mathcal{O}$ satisfies the semi-graphoid axioms: For disjoint subsets $W,X,Y$ and $Z$ of $\mathcal{O}$ it holds
\[
\begin{array}{lrcrl}
   (1) &  X \independent Y\,|Z \quad \,& \quad \quad\quad  \Rightarrow \quad \quad \quad  & Y \independent X\,|Z \,\,\,\, & \hbox{(symmetry)} \\
 (2) & X \independent (Y,W)\,|Z \quad\,  & \quad\quad\quad \Rightarrow \quad\quad\quad & \left\{\begin{array}{ll}X\independent Y\,|Z\\X\independent W\,|Z\end{array}\right. & \hbox{(decomposition)}\\
(3) & X \independent (Y,W)\,|Z\quad\, & \quad \quad\quad\Rightarrow \quad \quad\quad & X\independent Y | (Z,W)\,\,\,\,& \hbox{(weak union)}\\
(4) & \left.\begin{array}{r} X\independent W\,|(Z,Y) \\ X\independent Y\,|Z
    \end{array}\right\} & \quad\quad\quad \Rightarrow\quad\quad\quad  & X\independent (W,Y)\,|Z \,\,\,\,& \hbox{(contraction)}
\end{array}
\]
\end{Lem}
The proof is immediate using non-negativity and the chain rule of mutual information. In the probabilistic context, the axiomatic approach to conditional independence has been presented by Dawid\,\cite{dawidCI}. The above Lemma is important, since it implies that a DAG that fulfills the local Markov condition with respect to a set of objects is an efficient partial\footnote{In general there may hold additional conditional independence relations among the observations that are not implied by the local Markov condition together with the semi-graphoid axioms. In fact, it is well known that there so called non-graphical probability distributions whose conditional independence structure can not be completely represented by any DAG.} representation of the conditional independence structure among the observations. Namely, conditional independence relations can be read off the graph with the help of a criterion called d-separation\,\cite{pearlCausality}. This is the content of the following theorem but before stating it we recall the definition of d-separation: Two sets of nodes $A$ and $B$ of a DAG are d-separated given a set $C$ disjoint from $A$ and $B$ if every undirected path between $A$ and $B$ is blocked by $C$. A path that is described by the ordered tuple of nodes $(x_1,x_2,\ldots,x_r)$ with $x_1\in A$ and $x_r\in B$ is blocked if at least one of the following is true
\begin{enumerate}
\item[$(1)$] there is an $i$ such that  $x_i\in C$  and $x_{i-1}\rightarrow x_i \rightarrow x_{i+1}\;$ or $\;x_{i-1}\leftarrow x_i \leftarrow x_{i+1}\;$  or $\;x_{i-1}\leftarrow x_i \rightarrow x_{i+1}$\,,
\item[$(2)$] there is an $i$ such that $x_i$ and its descendants are not in $C$ and $x_{i-1}\rightarrow x_i \leftarrow x_{i+1}$.
\end{enumerate}
\begin{Thm}[Equivalence of Markov conditions]$ $\label{thmDec}\\
Let $I$ be a measure of mutual information on elements $O_{[n]}=\{O_1,\ldots,O_n\}$  and let $G$ be a DAG with node set $O_{[n]}$. Then the following two properties are equivalent
\begin{enumerate}
\item[$(1)$] (local Markov condition) Every node $O_i$ of $G$ is independent of its non-descendants $O_{nd}$ given its parents $O_{pa_i}$,
$$
O_i \independent O_{nd_i}\,|O_{pa_i}\,.
$$
\item[$(2)$] (global Markov condition) For every three disjoint sets of nodes $A$, $B$ and $C$ such that $A$ is $d$-separated from $B$ given $C$ in $G$, it holds $A \independent B\,|C$.
\end{enumerate}
\end{Thm}
Proof: $(1)\rightarrow (2)$. Since the dependence measure $I$ satisfies the semi-graphoid axioms (Lemma \ref{lemGraphoid}) we can apply Theorem 2 in Verma \& Pearl \cite{vermaPearl90} which asserts that the DAG is an $I$-map, or in other words that d-separation relations represent a subset of the (conditional) independences that hold for the given objects.\\
$(2)\rightarrow (1)$ holds because the non-descendants of a node are d-separated from the node itself by the parents. $\Box$\\
\section{Proof of Proposition \ref{propInd}}\label{propIndProof}
We have shown in Lemma \ref{lemSubMod} the submodularity of $I(Y:\cdot)$ with respect to independent sets. The rest of the proof is on the lines of the proof of Corollary I in \cite{madiman08}: First, by iteratively applying the chain rule for mutual information we obtain
\begin{equation}\label{chainRule}
I(Y:X_{[r]}) = \sum_{i=0}^{r-1} I(Y:X_{i+1}|X_{[i]}).
\end{equation}
Without loss of generality we can assume that every $X_i$ is part of at least one set $O_k$ for some $k$. Let $n_i$ be the total number of subsets $O_k$ containing $X_i$. By definition of $d_k$, for every $k$ it holds $n_i \leq d_k$ and we obtain
\begin{equation}\label{covpack}
\sum_{O_j,\, (X_i\in O_j)} \frac{1}{d_j} \leq n_i \cdot \max_{O_j,(X_i\in O_j)}{\frac{1}{d_j}} \leq  1\,.
\end{equation}
Putting (\ref{chainRule}) and (\ref{covpack}) together we get
\begin{eqnarray*}
I(Y:O_{[n]}) &=& I(Y:X_{[r]}) = \sum_{i=0}^{r-1} I(Y:X_{i}|X_{[i-1]})\\
 &\geq &  \sum_{i=1}^{n} I(Y:X_{i}|X_{[i-1]})\big(\sum_{O_j,\, (X_i\in O_j)} \frac{1}{d_j}\big)\,\\
&\stackrel{(a)}{=}&  \sum_{j=1}^n  \frac{1}{d_j} \sum_{X_i\in O_j} I(Y:X_i|X_{[i-1]})\\
&\stackrel{(b)}{\geq} & \sum_{j=1}^n  \frac{1}{d_j} \sum_{X_i\in O_j} I(Y:X_i|X_{[i-1]}\cap O_j)\\
&\stackrel{(c)}{=} & \sum_{j=1}^n  \frac{1}{d_j} I(Y:O_j)\,,
\end{eqnarray*}
where $(a)$ is obtained by exchanging summations and $(b)$ uses the property of $I$, that conditioning on independent objects can only increase mutual information (inequality (\ref{ineqCond}) applied to  $X_i \independent (X_{[i-1]}\backslash O_j)\,|O_j$)\,. This is the point at which submodularity of $I$ is used, since it is actually equivalent to (\ref{ineqCond}) as can be seen from the proof of Lemma \ref{lemSubMod}. Finally $(c)$ is an application of the chain rule to the elements of each $O_j$ separately. 

\section{Proof of Theorem \ref{thmAn}}\label{thmAnProof}
By  assumption $O_{i}\subseteq \mathcal{X}$ and the DAG $G$ with node set $\mathcal{X}$ fulfills the local Markov condition. For each $O_i$ denote by $an_G(O_i)$ the smallest ancestral set in $G$ containing $O_i$.\\
An easy observation that we need in the proof is given by the fact that two ancestral sets $A$ and $B$ are independent given their intersection:
\begin{equation}\label{eqCentAn}
A\backslash B\;\;\independent B\backslash A\;\;\;| A\cap B\,.
\end{equation}
This is implied by d-separation using Theorem \ref{thmDec}.\\
We first prove the inequality
\begin{equation}\label{eqthmProof}
I(Y:\,an_G(O_{[n]})) \geq \sum_{i=1}^{n} \frac{1}{d_i}I(Y:an_G(O_i))\,.
\end{equation}
From this the inequalities of the theorem follow directly: (\ref{ineqInd}) holds since $I(Y:an(O_i))\geq I(Y:O_i)$ using the monotony of $I$ (implied by chain rule and non-negativity). Further, (\ref{ineqInd2}) is a direct consequence of (\ref{eqthmProof}) together with the independence assumption $(\ref{indYY})$, since by the chain rule
$$
I(Y:\,an_G(O_{[n]})) = I(Y:O_{[n]}) + I(Y:\,an_G(O_{[n]})\backslash O_{[n]}\,|O_{[n]}) = I(Y:O_{[n]})\,,
$$
where the last equality is a consequence of $(\ref{indYY})$.\\
The proof of (\ref{eqthmProof}) is by induction on the number of elements in $\mathcal{A} = an_G(O_{[n]})$. If $\mathcal{A}=\emptyset$ nothing has to be proven. Assume now (\ref{eqthmProof}) holds for $\tilde{O}_{[n]}=\{\tilde{O}_1,\ldots,\tilde{O}_n\}$ such that $\tilde{\mathcal{A}}=\cup_{i=1}^n an(\tilde{O}_i)$ is of cardinality at most $k-1$. Let $O_{[n]}$ be a set of observations such that $\mathcal{A}$ is of cardinality $k$. From $O_{[n]}$ we construct a new collection $\tilde{O}_{[n]}$ as follows: W.l.o.g. assume $m:=d_1>0$, in particular $O_1$ is non-empty and moreover, by definition of $d_1$ and after reordering of the $O_i$ we can assume that the intersection $V := \cap_{i=1}^{m} an_G(O_{i})$ is non-empty. Note that $V$ itself is an ancestral set. We define $\tilde{O}_i = O_i\backslash V$ for all $1\leq i \leq n$ and denote by $\tilde{G}$ the modified graph that is obtained from $G$ by removing all elements of $V$. Further, denote by  $\tilde{I}(A:B\,|C) := I(A:B\,|C,V)$ a modified measure of mutual information obtained by conditioning on $V$. One checks easily that the graph $\tilde{G}$ fulfills the local Markov condition with respect to the independence relation induced by $\tilde{I}$ and is a DAG-model of the elements $\tilde{O}_{[n]}$. Hence, by induction assumption
\begin{equation}\label{eqthm1}
\tilde{I}\big(Y:an_{\tilde{G}}(\tilde{O}_{[n]})\big) \geq \sum_{i=1}^n \frac{1}{\tilde{d}_i} \tilde{I}\big(Y:an_{\tilde{G}}(\tilde{O}_{i})\big)\,,
\end{equation}
where $\tilde{d}_i$ is defined similarly as $d_i$, but with respect to the elements $\tilde{O}_i$ and $\tilde{G}$. Further the sum is over all non-empty $\tilde{O}_i$. By construction of $\tilde{I}$ and $\tilde{O}_{[n]}$, the left hand side of  (\ref{eqthm1}) is equal to
\begin{eqnarray}
\tilde{I}\big(Y:an_{\tilde{G}}(\tilde{O}_{[n]})\big)  &=& I\big(Y:an_G(O_{[n]})\backslash V\;|V\big) = I(Y:an_G(O_{[n]})) - I(Y:V)\,\label{eqThm2}\,.
\end{eqnarray}
The right hand side of (\ref{eqthm1}) can be rewritten to
\begin{eqnarray*}
\sum_{i=1}^n \frac{1}{\tilde{d}_i} \tilde{I}\big(Y:an_{\tilde{G}}(\tilde{O}_{i})\big) &\stackrel{(a)}{\geq}& \sum_{i=1}^n \frac{1}{d_i} \tilde{I}\big(Y:an_{\tilde{G}}(\tilde{O}_{i})\big)\\
&\stackrel{(b)}{=} &
\sum_{i=1}^m \frac{1}{d_i} I(Y:an_G(O_i)\backslash V\,|V) + \sum_{i=m+1}^n \frac{1}{d_i} I(Y:an_G(O_i)\,|V)\,\\
&\stackrel{(c)}{\geq} &
\sum_{i=1}^m \frac{1}{d_i} I(Y:an_G(O_i)\backslash V\,|V) + \sum_{i=m+1}^n \frac{1}{d_i} I(Y:an_G(O_i))\,,\\
\end{eqnarray*}
where $(a)$ follows because $d_i\geq \tilde{d}_i$ by definition and $(b)$ follows because $an_G(O_i)\cap V =\emptyset$ for $i>m$. Hence by (\ref{eqCentAn}) $V$ and $an_G(O_i)$ are independent and therefore conditioning on $V$ only increases mutual information as proven in Lemma \ref{lemPropMut} and inequality $(c)$ follows. We continue by rewriting the first $m$ summands of the right hand side using the chain rule
\begin{eqnarray*}
\sum_{i=1}^m \frac{1}{d_i} I(Y:an_G(O_i)\backslash V\,|V) &=& \sum_{i=1}^m \frac{1}{d_i} \big[ I(Y:an_G(O_i)) - I(Y:V) \big]\\
&\geq & \Big[\sum_{i=1}^m \frac{1}{d_i} I(Y:an_G(O_i))\Big] - I(Y:V)\,,
\end{eqnarray*}
where the inequality holds because $\sum_{i=1}^m \frac{1}{d_i} \leq 1$ which has already been used, see (\ref{covpack}) in the proof of Proposition \ref{propInd}\,. Summarizing, the right hand side of (\ref{eqthm1}) can be bounded from below by
$$
\sum_{i=1}^n \frac{1}{\tilde{d}_i} \tilde{I}(Y:an_{\tilde{G}}(\tilde{O}_{i})) \geq   \sum_{i=1}^n \frac{1}{d_i}I(Y:an_G(O_i)) - I(Y:V)\,.
$$
Since we have shown in $(\ref{eqthm1})$ and $(\ref{eqThm2})$, that the left hand side can be bounded from above by $I(Y:O_{[n]})-I(Y:V)$, we observe that $I(Y:V)$ cancels and (\ref{eqthmProof}) is proven.

\section{Proof of Corollary \ref{corLoc}}\label{apCorProof}

Proof: 
Let $G$ be a DAG-model of the observation of $O_{[n]}=\{O_1,\ldots,O_n\}$. We construct a new DAG $G'$, by removing the objects of $A :=  \cup_{i=1}^n A_{c_i+1}$. Since $A$ is an ancestral set $G'$ fulfills the local Markov condition with respect to the mutual information measure obtained by conditioning on $A$. We apply Theorem \ref{thmAn} to $G'$ and the observations $O'_{[n]} = \{O_1\backslash A,\ldots,O_n\backslash A\}$ to get 
\begin{equation}\label{ineqCor2}
I(Y : an_{G'}(O'_{[n]})\,|A) \geq \sum_{i=1}^n \frac{1}{c_i} I(Y:O'_i\,|A)\,.
\end{equation}
Using assumption $(\ref{ineqCor})$ and the chain rule for mutual information we obtain
\begin{eqnarray*}
I(Y:A) &=& I(Y:an_G(O_{[n]})) - I(Y:an_G(O_{[n]})\backslash A\,| A)\\
&\stackrel{(a)}{=}& I(Y:an_G(O_{[n]})) - I(Y:an_{G'}(O'_{[n]})\,| A)\\
&\stackrel{(b)}{\leq}& \sum_{i=1}^{n} \frac{1}{c_i}\big[I(Y:O_i)-I(Y:O'_i|A)\big] - \epsilon_{\mathbf{c}}\\
&\stackrel{(c)}{\leq} & \sum_{i=1}^{n} \frac{1}{c_i}I(Y:A) - \epsilon_{\mathbf{c}}\,,
\end{eqnarray*}
where in $(a)$ we used the definition of $O'_i$ and for $(b)$ we plugged in inequalities (\ref{ineqCor}) and (\ref{ineqCor2}). Finally $(c)$ holds because
\begin{eqnarray*}
I(Y:O_i)-I(Y:O_i'|A) &=& I(Y:O_i\cap A|O_i') + I(Y:O_i') - I(Y:O_i'|A)\\
&=& I(Y:O_i\cap A|O_i') + I(Y:A) - I(Y:A|O_i') \leq I(Y:A)\,,
\end{eqnarray*}
where the chain rule has been applied multiple times. The corollary now follows by solving for $I(Y:A)$.
$\Box$

%\bibliographystyle{ieeetr}
%\bibliography{../referencesPHD}

\begin{thebibliography}{10}

\bibitem{pearlCausality}
J.~Pearl, {\em Causality}.
\newblock Cambridge University Press, 2000.

\bibitem{spirtes}
P.~Spirtes, C.~Glymour, and R.~Scheines, {\em Causation, Prediction, and
  Search, Second Edition (Adaptive Computation and Machine Learning)}.
\newblock The MIT Press, 2001.

\bibitem{lauritzen96}
S.~L. Lauritzen, {\em Graphical Models}.
\newblock Oxford Statistical Science Series, {Oxford University Press, USA},
  July 1996.

\bibitem{dominikK}
D.~Janzing and B.~Sch\"olkopf, ``Causal inference using the algorithmic markov
  condition,'' {\em IEEE Trans. Inf.Th.}, vol.~56, oct. 2010.

\bibitem{bastianCOLT}
B.~Steudel, D.~Janzing, and B.~Sch{\"o}lkopf, ``Causal markov condition for
  submodular information measures,'' {\em Proceedings of COLT 2010},
  vol.~abs/1002.4020, 2010.

\bibitem{reichenbach}
H.~Reichenbach, {\em The Direction of Time}.
\newblock University of Califonia Press, 1956.

\bibitem{coverThomas}
T.~M. Cover and J.~A. Thomas, {\em Elements of Information Theory}.
\newblock Wiley-Interscience, 2nd~ed., July 2006.

\bibitem{algStat}
P.~G\'{a}cs, J.~T. Tromp, and P.~M. Vit\'{a}nyi, ``Algorithmic statistics,''
  {\em IEEE Trans. Inf.Th.}, vol.~47, pp.~2443--2463, 2001.

\bibitem{pearlBelief}
J.~Pearl, {\em Probabilistic reasoning in intelligent systems: networks of
  plausible inference}.
\newblock San Francisco, CA, USA: Morgan Kaufmann Publishers Inc., 1988.

\bibitem{infoCausQuant}
M.~{Paw{\l}owski}, T.~{Paterek}, D.~{Kaszlikowski}, V.~{Scarani}, A.~{Winter},
  and M.~{{\.Z}ukowski}, ``{Information causality as a physical principle},''
  {\em Nature}, vol.~461, pp.~1101--1104, Oct. 2009.

\bibitem{dawidCI}
A.~P. Dawid, ``Conditional independence in statistical theory,'' {\em Journal
  of the Royal Statistical Society. Series B (Methodological)}, vol.~41, no.~1,
  pp.~1--31, 1979.

\bibitem{madiman08}
M.~Madiman and P.~Tetali, ``Information inequalities for joint distributions,
  with interpretations and applications,'' {\em IEEE Trans. Inf.Th.}, 2008.

\bibitem{stillNetworkInfo}
E.~Schneidman, S.~Still, M.~J.~B. II, and W.~Bialek, ``Network information and
  connected correlations,'' {\em Phys. Rev. Let.}, vol.~91, 2003.

\bibitem{studenyMulti}
M.~Studeny and J.~Vejnarova, ``The multiinformation function as a tool for
  measuring stochastic dependence,'' {\em M. I. Jordan (ed), Learning in
  Graphical Models}, pp.~261--297, 1998.

\bibitem{speed}
J.~N. Darroch, S.~L. Lauritzen, and T.~P. Speed, ``Markov fields and log-linear
  interaction models for contingency tables,'' {\em Annals of Statistics},
  vol.~8, pp.~522--539, 1980.

\bibitem{amari}
S.~I. Amari, ``Information geometry on hierarchy of probability
  distributions,'' {\em IEEE Trans. Inf.Th.}, vol.~47, no.~5, pp.~1701--1711,
  2001.

\bibitem{bookKC}
M.~Li and P.~Vit\'{a}nyi, {\em An Introduction to Kolmogorov Complexity and Its
  Applications}.
\newblock Text and Monographs in Computer Science, Springer-Verlag, 2007.

\bibitem{pearlTestability}
J.~Pearl, ``On the testability of causal models with latent and instrumental
  variables,'' in {\em UAI}, pp.~435--443, 1995.

\bibitem{sullivantGauss}
S.~Sullivant, K.~Talaska, and J.~Draisma, ``Trek separation for gaussian
  graphical models,'' {\em arXiv:0812.1938}, 2009.

\bibitem{ricco07}
E.~Riccomagno and J.~Q. Smith, ``Algebraic causality: Bayes nets and beyond,''
  {\em http://arxiv.org/abs/0709.3377}, Sep 2007.

\bibitem{nihat2}
N.~Ay, ``A refinement of the common cause principle,'' {\em Discrete Applied
  Mathematics}, vol.~157, pp.~2439--2457, 2009.

\bibitem{vermaPearl90}
T.~Verma and J.~Pearl, ``Causal networks: Semantics and expressiveness,'' {\em
  Uncertainty in Artificial Intelligence 4}, pp.~69--76, 1990.

\end{thebibliography}

\end{document}